\documentclass[12pt,preprint]{aastex}

\slugcomment{To appear in ApJ Letters}

\shorttitle{Widespread HCO emission in M\,82}
\shortauthors{Garc\'{\i}a-Burillo et al.}

\begin{document}


\title{Widespread HCO emission in the M\,82's nuclear starburst 
\footnote{Based on observations carried out with the IRAM Plateau de Bure 
Interferometer. IRAM is supported by INSU/CNRS (France), MPG (Germany) and 
IGN (Spain).}}


\author{S.Garc\'{\i}a-Burillo \\
	J.Mart\'{\i}n-Pintado \\
	A.Fuente \\
	A.Usero}

\affil{Observatorio Astron\'omico Nacional (OAN), Campus Universitario \\ Apdo.1143, 
Alcal\'a de Henares, E-28800, Madrid, SPAIN}

\email{burillo@oan.es, martin@oan.es, fuente@oan.es, antonio.u@imaff.cfmac.csic.es}


\and

\author{R. Neri}
\affil{Institut de Radio Astronomie Millim\'etrique (IRAM) \\
300, Rue de la Piscine, 38406-St.Mt.d'H\`eres, FRANCE}
\email{neri@iram.fr}

\begin{abstract}

We present a high-resolution ($\sim$5'') image of the nucleus of M\,82 showing 
the presence of widespread emission of the formyl radical (HCO). 
The HCO map, the first obtained in an external galaxy, reveals the existence of 
a structured disk of $\sim$650\,pc full diameter. The HCO distribution in the plane 
mimics the ring morphology displayed by other molecular/ionized gas tracers in M\,82. 
More precisely, rings traced by HCO, CO and HII regions are nested, 
with the HCO ring lying in the outer edge of the molecular torus. 
Observations of HCO in galactic clouds indicate that 
the abundance of HCO is strongly enhanced in the interfaces between the 
ionized and molecular gas. The surprisingly high overall abundance of HCO 
measured in M\,82 (X(HCO)$\sim$4\,10$^{-10}$) indicates that its nuclear disk 
can be viewed as a {\it giant} Photon Dominated Region (PDR)
of $\sim$650\,pc size. The existence of various nested gas rings, with the 
highest HCO abundance  occurring at the outer ring (X(HCO)$\sim$0.8\,10$^{-9}$), 
suggests that PDR chemistry is {\it propagating} in the disk. 
We discuss the inferred large abundances of HCO in M\,82 in the context of a 
starburst evolutionary scenario, picturing the M\,82 nucleus 
as an evolved starburst.

\end{abstract}


\section{Introduction}

M\,82 is the closest galaxy experiencing a massive star formation episode 
\citep{rie80, wil99}. Its nuclear starburst, located in the central 1\,kpc, 
has been the subject of continuum and line observations made in virtually all 
wavelengths from X-rays to the radio domain. These studies indicate that 
the high rate of supernova explosions and the strong UV radiation fields have heavily 
influenced the physical properties and kinematics of the interstellar medium in M\,82. The high supernova rate 
has created a biconical outflow of hot gas (Bregman, Schulman, \& Tomisaka 1995; 
Shopbell \& Bland-Hawthorn 1998) also observed 
in the cold gas and dust (Alton, Davis, \& Bianchi 1999; Seaquist \& Clark 2001). 
The discovery of a $\sim$500\,pc molecular gas chimney and a giant supershell in M\,82, 
detected in SiO, indicates the occurrence of large-scale shocks in the disk-halo interface of 
the starburst \citep{gbu01}. Furthermore, there are evidences that the  
strong UV-fields have created a particular physical environment in the molecular gas 
reservoir of M\,82 \citep{stu97,mao00,wei01}. The common picture emerging from these 
studies is that the bulk of CO emission in the nuclear disk of M\,82 comes 
from moderately dense (with n(H$_2$)$\sim$10$^{3}$-10$^{4}$cm$^{-3}$) Photon Dominated Regions (PDR). 
However, these conclusions are model-dependent and not free from 
internal inconsistencies \citep{mao00}. 

Observational evidence supports that the emission of the formyl radical (HCO) mainly arises from PDR at the interfaces between the 
ionized and the molecular gas in our Galaxy. 
After the first detection of HCO  by Snyder, Hollis, \& Ulich (1976)  
ulterior searches have confirmed that this radical is 
associated with regions where chemistry is driven by an enhanced UV radiation field 
(Hollis \& Churchwell 1983; Snyder, Schenewerk, \& Hollis 1985; Schenewerk et al.1988; 
Schilke et al. 2001). On the theoretical side, several chemical models have been put forward to account for 
the observed abundances of HCO in PDR (de Jong, Boland, \& Dalgarno 1980; 
Leung, Herbst, \& Huebner 1984; Schilke et al. 2001).

We have chosen HCO, a privileged tracer of PDR chemistry, 
to investigate the influence of the M\,82 starburst on its molecular gas 
reservoir. \citet{sag95} tentatively detected the emission of HCO at 3mm in M\,82, using 
single-dish observations. However, the HCO lines appear blended with SiO 
and H$^{13}$CO$^{+}$ lines in their spectrum. Furthermore, the low spatial 
resolution (50'') of their single-pointed map did not allow to infer the spatial 
distribution of HCO. In this Letter we present a high-resolution ($\sim$5'') image of the emission
of HCO in the nucleus of M\,82. 
The interferometer HCO map of M\,82, the first obtained in an 
external galaxy, shows unambiguous evidence that the whole nuclear disk has 
become a {\it giant} PDR of $\sim$650\,pc size with a total 
HCO abundance of $\sim$4\,10$^{-10}$. The enhancement of 
HCO presents spatial variations which depend on the distance to
the most prominent HII regions of the M\,82's starburst.

\section{Observations}

The HCO observations of M\,82 were carried out with the IRAM array at
Plateau de Bure (France) during 1999 June, using the 4-antennas CD set of configurations. 
We observed four of the strongest N$_{K-K+}$=1$_{01}$-0$_{00}$ hyperfine lines of HCO. 
These transitions form a 3mm quartet: J=3/2--1/2, F=2--1(86.671\,GHz) 
and F=1--0(86.708\,GHz); J=1/2--1/2, F=1--1(86.777\,GHz) and F=0--1(86.806\,GHz). 
We observed simultaneously the J=2--1 line of SiO(86.847\,GHz), and the J=1--0 line of 
H$^{13}$CO$^+$(86.754\,GHz). The one-field observations 
are made centering the 55$''$ primary beam of the array at 
$\alpha_{J2000}$=$09^h55^m51.9^s$ and $\delta_{J2000}$=$69^{\circ}40'47.1''$, which 
corresponds to the 2.2$\mu$ peak (Joy, Lester, \& Harvey 1987). Calibration of visibilities 
was performed as explained in \citet{gbu01}. 
The synthesized beam is almost circular ($5.9''\times 5.6''$, PA=105$^{\circ}$). The rms noise level 
in 2.5\,MHz wide channel maps, derived after subtraction of the 
continuum emission, is 1\,mJy/beam (5\,mK). We assume a distance to M\,82 of D=3.9\,Mpc \citep{sak99}; 
the latter implies 1''$\sim$20\,pc. We define the $x$ and $y$ axes to run parallel to 
the galaxy major and minor axes, respectively, where $x>$0 eastward and $y>$0 northward. 
The major axis has PA=70$^{\circ}$; (x,y) offsets are referred to the 2.2$\mu$ peak.
 All through the paper, line intensities are not corrected from primary beam attenuation. 
Note however that the bulk of the discussion is based on the study of the ratios of 
lines which are equally affected by this bias. If they were to be 
corrected, HCO and H$^{13}$CO$^+$ intensities should be both 
multiplied by a factor $\sim$1.3 at the edge of 
the M\,82's nuclear disk  (at a$\sim$17'' radius).

\section{The HCO map}

Figure 1a represents the velocity-integrated intensity map obtained for the strongest hyperfine 
component (F=2--1) of the J=3/2--1/2 line of HCO. This image
shows that the emission of HCO is widespread in the nuclear starburst of M\,82. 
The bulk of the emission comes from a highly-structured disk of $\sim$650\,pc full diameter and barely 
resolved vertical thickness. The HCO distribution in the plane mimics the ring morphology displayed 
by other molecular gas tracers in M\,82 \citep{mao00}. 
Two emission peaks at (x,y)=(+16'',-2'') and (-15'',0'') locate the eastern and western 
lobes of the HCO ring viewed edge-on. A central clump at (+2'',0'') 
is close to the 2.2$\mu$ peak identified as the dynamical center \citep{gbu01}. 
With lower statistical significance, we detect out-of-the plane 
emission in a clump at (-9'',8''). However, the full-size of this structure 
is smaller than the beam, and therefore its reality is doubtful. 

Figure 1b shows the spectra of HCO, SiO and H$^{13}$CO$^{+}$ integrated in the regions of
the two ring maxima (see Figure 1a). Gaussian profiles are fitted for the 
identified lines. 
Together with the strongest F=2--1 line, emission of the F=1--0 component is detected at 
4$\sigma$ level in the two ring lobes. We have also 
detected the F=1--1 and F=0--1 transitions in the western and eastern lobes, respectively. 
We estimate an average F=2--1/F=1--0 line ratio of $\sim$3$\pm$1 ($\sim$1.5$\pm$0.5)
for the western (eastern) lobe. 
Within the errors, these ratios are close to the theoretical line 
ratio of $\sim$1.7, expected for optically thin emission.

Figures 2a-b show the HCO(F=2--1) integrated intensity contours overlaid  with 
the zero moment maps of H$^{13}$CO$^{+}$ (1--0) \citep{gbu01} 
and $^{12}$CO(2--1) \citep{wei01}. Although all images reveal a ring-like 
distribution of molecular gas in the nucleus of M\,82, the separation between the E/W 
lobes differ significantly between HCO ($\sim$32'') and both $^{12}$CO and H$^{13}$CO$^{+}$ 
($\sim$25''). The HCO emission extends farther out in the disk, especially in the eastern lobe.
The HCO torus encircles the molecular torus traced by H$^{13}$CO$^{+}$ and CO. Most remarkably, 
the ring of HII regions (of 10''-15'' size) identified by Br$\gamma$ 
\citep{lar94}, NeII \citep{ach95}, and radio-recombination lines (e.g. the H41$\alpha$ 
map of Seaquist et al. 1996) lies just inside the CO molecular ring. The HCO, CO and HII rings are nested 
with HCO lying in the outer edge of the disk. 
The central HCO maximum is also displaced from the $^{12}$CO/H$^{13}$CO$^{+}$ hot spot 
at (-3'',0''). At smaller scales also, the peaks of HCO appear to 
avoid the brightest HII regions (see section 4). To interpret the described 
morphology we will first estimate the spatial distribution of the abundance of HCO 
in M\,82 and discuss different scenarios accounting for it.
 
Kinematics of HCO gas in M\,82's disk show no significant departures from the general rotation 
pattern typically displayed by other molecular gas tracers in this galaxy.  
   
\section{The HCO abundance map in M\,82}

Figure 3 represents in gray scale the HCO(F=2--1)/H$^{13}$CO$^{+}$(1--0) 
intensity ratio map in the nucleus of M\,82, derived using a 6$\sigma$ (4$\sigma$) clipping 
for H$^{13}$CO$^{+}$(1--0) (HCO(F=2--1)). The inferred ratio, accurate to 30$\%$ 
varies between $\sim$0.50$\pm$0.15 (at the outer edges of the ring, and in the central peak) 
to $\sim$0.15$\pm$0.04 (in the region between the HCO maxima), with an average
ratio of $\sim$0.25$\pm$0.07 for the whole disk. In order to derive the HCO to H$^{13}$CO$^{+}$ 
column density ratio we need to make several assumptions about both the optical thickness 
and excitation temperature of the lines. Based on studies of 
HCO emission in galactic PDR \citep{sch86, sch88} it is plausible 
to suppose that the HCO lines should be optically thin also
in M\,82. The hyperfine line ratios measured in M\,82 seem to confirm these expectations.
As excitation temperature for HCO we will assume $T_{\rm ex}$=10$K$; this value was derived
from a multitransition analysis made by \citet{sny85} on NGC\,2024.

For H$^{13}$CO$^{+}$ we will also consider optically thin emission \citep{gbu00} and the same
excitation temperature as assumed for HCO. These are reasonable guesses, especially for 
$T_{\rm ex}$, as the two molecules have similar critical densities for the examined 
transitions. In this case the calculation of HCO to H$^{13}$CO$^{+}$ column density ratio is straightforward 
using the expression \citep{sch88}:

\begin{equation}
\frac{N(HCO)}{N(H^{13}CO^{+})} \simeq \frac{12}{g_{U}}\,\frac{I_{HCO}\times A^{-1}_{HCO}}{I_{H^{13}CO^{+}}\times A^{-1}_{H^{13}CO^{+}}}
\end{equation}

where $N$ is the total column density, $g_{U}$ is the degeneracy of the upper hyperfine level 
($g_{U}$=5 for $F$=2), $I$ is the integrated intensity, and $A$ is the Einstein 
coefficient of the transition. We derive values for $N(HCO)/N(H^{13}CO^{+})$ that 
range from 2 to 8, with an average value of 3.6 for the whole disk. 
Adopting an average fractional abundance for H$^{13}$CO$^{+}$ of 
$\sim$10$^{-10}$ \citep{gbu00,gbu01}, we derive X(HCO)$\sim$4\,10$^{-10}$ in M\,82. 
   
\citet{sch88} measured a value of $N(HCO)/N(H^{13}CO^{+})\sim$9.7 in the 
HII region NGC\,2024. On the contrary, values of $N(HCO)/N(H^{13}CO^{+})$ 
significantly lower than 1, i.e. an order magnitude below those found in 
NGC\,2024, have been reported by \citet{sch88} in galactic clouds without 
developed HII regions or having no indication 
of star formation. More recently, \citet{sch01} have searched for 
HCO in a reduced sample of prototypical PDR. 
The estimated $N(HCO)/N(H^{13}CO^{+})$ abundance 
ratios range from $\sim$30 (in the Orion bar) to $\sim$3 (in NGC\,7023). 
The largest HCO abundances are found in the Orion Bar: the paradigm of
interaction between an HII region (M\,42) and its parent molecular cloud.
Comparing our global estimates of $N(HCO)/N(H^{13}CO^{+})\sim$3.6 in M\,82 
with the values measured for the prototypical PDR in our Galaxy, we conclude that the 
nuclear disk of M\,82 can be viewed as a {\it giant} PDR. 
Further analysis of the spatial variation of the $N(HCO)/N(H^{13}CO^{+})$ ratio shows
that the chemistry driven by the strong UV fields 
has produced the highest HCO enhancement in the outer edges of the starburst ring; 
the central clump also shows a similar HCO enrichment ($N(HCO)/N(H^{13}CO^{+})\sim$7--8). 
These regions, with HCO abundances approaching those of NGC\,2024, 
lie close to the loci of HII regions in M\,82 but do not quite coincide
spatially with them. A mere inspection of Figure 3 indicates that the HCO-enriched molecular 
clouds surround the ionized gas as shown by the strongest [NeII] emission: X(HCO)/[NeII] 
peaks avoid each other and are shifted by offsets of $\sim$50-150pc. This might seem paradoxical 
as a close correlation between HII regions and PDR could be expected at these scales.  
As discussed below, this distribution could be explained however as the final result of 
{\it propagation} of PDR chemistry in M\,82.

\section{Discussion}

The enhancement of HCO in PDR is mostly an observational fact, 
reproduced with uneven success by models. PDR models published so far have tried to account  
for the derived abundance of HCO using either gas-phase schemes 
\citep{jon80,leu84} or incorporating dust grains to the chemistry \citep{sch01}.  
According to \citet{jon80}, the C$^+$ ion, highly abundant in UV-processed 
cloud envelopes, would start the chain of reactions leading to CH$_2$ and 
finally, to HCO. Confirming these expectations, \citet{hol83} found an empirical 
correlation between HCO and C$^+$ radio recombination line emission in two 
galactic clouds. \citet{sch01} have proposed a model where HCO is the 
final by-product of the photodesorption or evaporation of solid formaldehyde 
on dust grain mantles \citep{wes95}.  Although the values of X(HCO) predicted by 
\citet{sch01} are still one order of magnitude below what is typically found in PDR, 
these results are encouraging. Based on a 1-dimensional model made for the 
Orion molecular cloud, \citet{sch01} predict that the large photodissociation rate 
of HCO \citep{dis88} could be counterbalanced only for extinctions A$_v>$5-6. 
For a population of clouds uniformly bathed in the pervasive UV 
field of a starburst galaxy, the A$_v$ limit should be  
risen by a factor of $\sim$a few, making the equivalent condition on column density 
close to N(H$_2$)$>$10$^{22}$cm$^{-2}$.

The HCO map of M\,82 provides direct evidence that the  
starburst event has heavily processed the bulk of molecular gas in this galaxy. 
The global fractional abundance of HCO ($\sim$4\,10$^{-10}$, averaged over $\sim$650\,pc) 
can be accounted within a PDR scenario. 
Paradoxically, the spatial correlation between 
HCO-enriched clouds and HII regions is poor however. This is noticeable both at large scales 
(HCO/CO/HII nested ring morphology) but also at small scales (peaks of HCO/HII emission 
avoid each other). The poor correlation suggests that UV fields coming from the 
strongest HII regions of M\,82, lying in the inner $\sim$400\,pc, have photodissociated 
the bulk of HCO in nearby molecular cloud envelopes. 
Previous observational evidence pointed out to a 
disrupted physical environment for the dense ISM in M\,82. 
Observations of the atomic carbon [CI]\,$^{3}$P$_{2}\rightarrow^{3}$P$_{1}$ and
 $^{3}$P$_{1}\rightarrow^{3}$P$_{0}$ lines \citep{sch93,stu97} confirmed that 
the M\,82 [CI]/CO abundance ratio ($\sim$0.5) is
higher than observed in non-starburst disks (it is $\sim$.15 in our Galaxy). 
Moreover, the measured J=2--1/1--0 ratio can only be reconciled with
clouds being small (with sizes $\sim$1\,pc), only moderately 
dense (with densities $\sim$10$^{3}$-10$^{4}$cm$^{-3}$), and hot (with temperatures $\sim$50-100K) 
\citep{stu97}. Similar conclusions, based on Large Velocity Gradient (LVG) and PDR 
modelling of $^{12}$CO, $^{13}$CO, and C$^{18}$O line ratios have been 
reported by \citet{mao00} and \citet{wei01}. The bulk of molecular clouds in the inner 
$\sim$400\,pc of M\,82 might have low densities (at some places in the disk as low as 
$\sim$10$^{3}$cm$^{-3}$) and small sizes ($\sim$1\,pc). Assuming this scenario, 
the survival of HCO against photodissociation, requiring column densities 
N(H$_2$)$>$10$^{22}$cm$^{-2}$ might be difficult.
Most remarkably, the location of HCO-enriched clouds in our map, identified by their 
largest $N(HCO)/N(H^{13}CO^{+})$ ratio (the outer ring and 
the central clump), coincides with Weiss et al.'s model predictions on where  
individual molecular clouds may have the largest column 
densities (where volume densities reach a few $\sim$10$^{4}$cm$^{-3}$). Furthermore, the 
intensity of UV fields is there comparatively lower (see Figure 3).

The reported differences between HCO and H$^{13}$CO$^{+}$ maps in M\,82 cannot be easily 
explained with the present PDR models however, as both species are expected to be closely 
associated in PDR. Other unexplored scenarios cannot be excluded as plausible 
explanations for the enhancement of HCO: X-Ray or Cosmic Ray Dominated Regions 
chemistries might account for it in the case of M\,82 (Suchkov, Allen, \& Heckman 1993). 
Moreover, we can not exclude the presence of dense molecular gas 
(with n(H$_2$)$>$10$^{5}$cm$^{-3}$) in the central 1\,kpc of M\,82 as already pointed 
out by \citet{mao00} and \citet{wei01}, or shown by the detection
of tracers of dense molecular gas (see HCN map of Brouillet \& Schilke 1993). 
However, our results suggest that evaporation/destruction of molecular clouds 
by UV fields is highly efficient in the inner $\sim$400\,pc of the galaxy. 
This efficiency shows spatial trends within the M\,82's disk but the reasons 
behind these variations remain to be understood.

M\,82 is the only galaxy where emission of HCO has been detected so far. 
HCO emission was also searched for in the starburst galaxy 
NGC\,253 by \citet{gbu00}. Their data, giving no detection, allowed to set
a 3-$\sigma$ upper limit for the  HCO/H$^{13}$CO$^{+}$  ratio of $\le$ 0.12, i.e. 
a factor of 4-5 smaller than the value measured in the M\,82's ring. 
HCO is the only {\it complex} molecule showing a larger fractional 
abundance in M\,82 than in NGC\,253 \citep{mau93}. 
Furthermore, the remarkably different abundances and spatial distributions 
of SiO gas in NGC\,253 and M\,82 are suggestive of an 
evolutionary link between these starbursts \citep{gbu00,gbu01}.
The large HCO abundance in M82 may fit within this scenario, 
which considers the M\,82 starburst episode as more evolved than in NGC\,253.
The chemistry of molecular gas in NGC\,253 is heavily influenced by the large-scale 
shocks and heating induced by a burst of pre-main sequence massive stars.
This explains the high abundance of SiO and 
the significantly low abundance of HCO. The chemistry of molecular gas in M\,82 
is dominated by the action of UV fields produced by more evolved massive stars 
giving rise to HII regions. The remnant of the starburst 
presents a PDR-like chemistry favoring the presence of HCO, but forcing a 
low abundance for other complex molecules.

\acknowledgments

This work received support from the Spanish DGES under grant number 
AYA, 2000-0927 and CICYT-PNIE under grant ESP2001-4519-PE. We acknowledge 
the IRAM staff for carrying the observations and help during data reduction.




\clearpage

  \figcaption{
The top panel (a) shows the velocity-integrated intensity map of HCO(F=2--1) in the
central region of M\,82 (contour levels: -0.09, 0.144 to 0.350Jy beam$^{-1}$km s$^{-1}$ 
by steps of 0.024Jy beam$^{-1}$km s$^{-1}$; $\sigma$=0.048Jy beam$^{-1}$km s$^{-1}$ ). 
A filled ellipse  on the lower right corner pictures the (5.9"x5.6") synthesized 
beam. The dashed line traces the galaxy major axis (PA=70$^{o}$). ($\Delta$X,$\Delta$Y) offsets, along the major and minor axis 
respectively, are referred to the dynamical center of the galaxy, marked
by the central filled square. The bottom pannel (b) shows the spectra integrated over the squared regions covering
the eastern and western lobes as shown. The hyperfine components of HCO detected 
are labeled by their "F-jump". H$^{13}$CO$^{+}$(J=1--0) and SiO (v=0, J=2--1) spectra 
are also plotted.\label{fig1}}

\figcaption{Top panel (a) shows the comparison of the integrated-intensity maps of 
HCO (F=2--1) (contours: 0.144 to 0.350Jy beam$^{-1}$km s$^{-1}$ by steps of 
0.024Jy beam$^{-1}$km s$^{-1}$) and H$^{13}$CO$^{+}$(J=1--0) 
(gray scale: 0.280-1 Jy beam$^{-1}$km s$^{-1}$). Bottom panel (b) shows
comparison of HCO (F=2--1) and $^{12}$CO(2--1) integrated-intensity maps 
(gray scale for CO: 200-550Jy beam$^{-1}$km s$^{-1}$).
\label{fig2}}

\figcaption{Map of the HCO(F=2--1)/H$^{13}$CO$^{+}$(J=1--0) integrated-intensity 
ratio (gray scale from 0.15 to 0.5) compared with the [NeII] flux map of \citet{ach95}, 
in contours linearly scaled from 10$\%$ to 90$\%$ by steps of 10$\%$ of the peak value .
\label{fig3}}

\clearpage 

\plotone{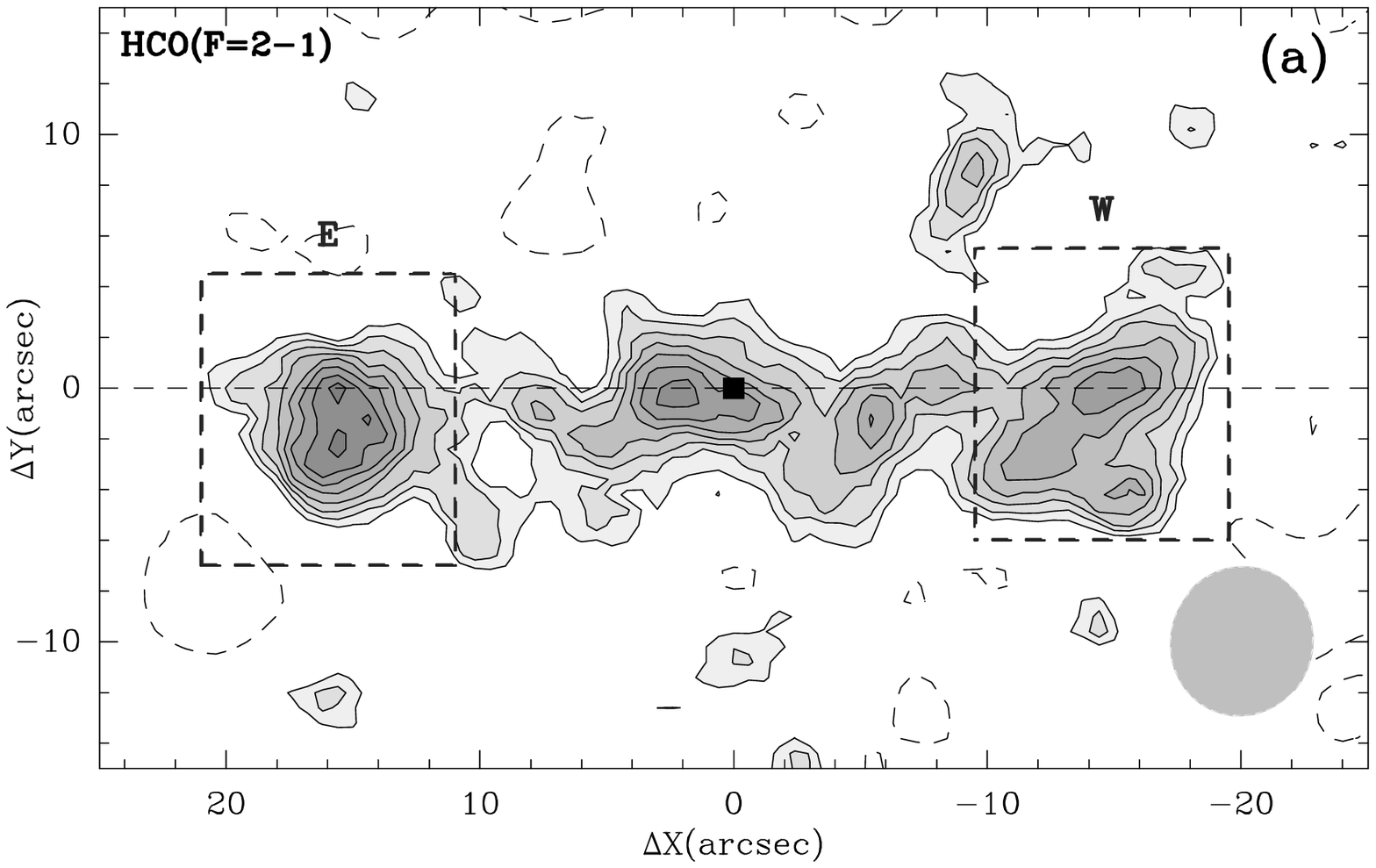}

\plotone{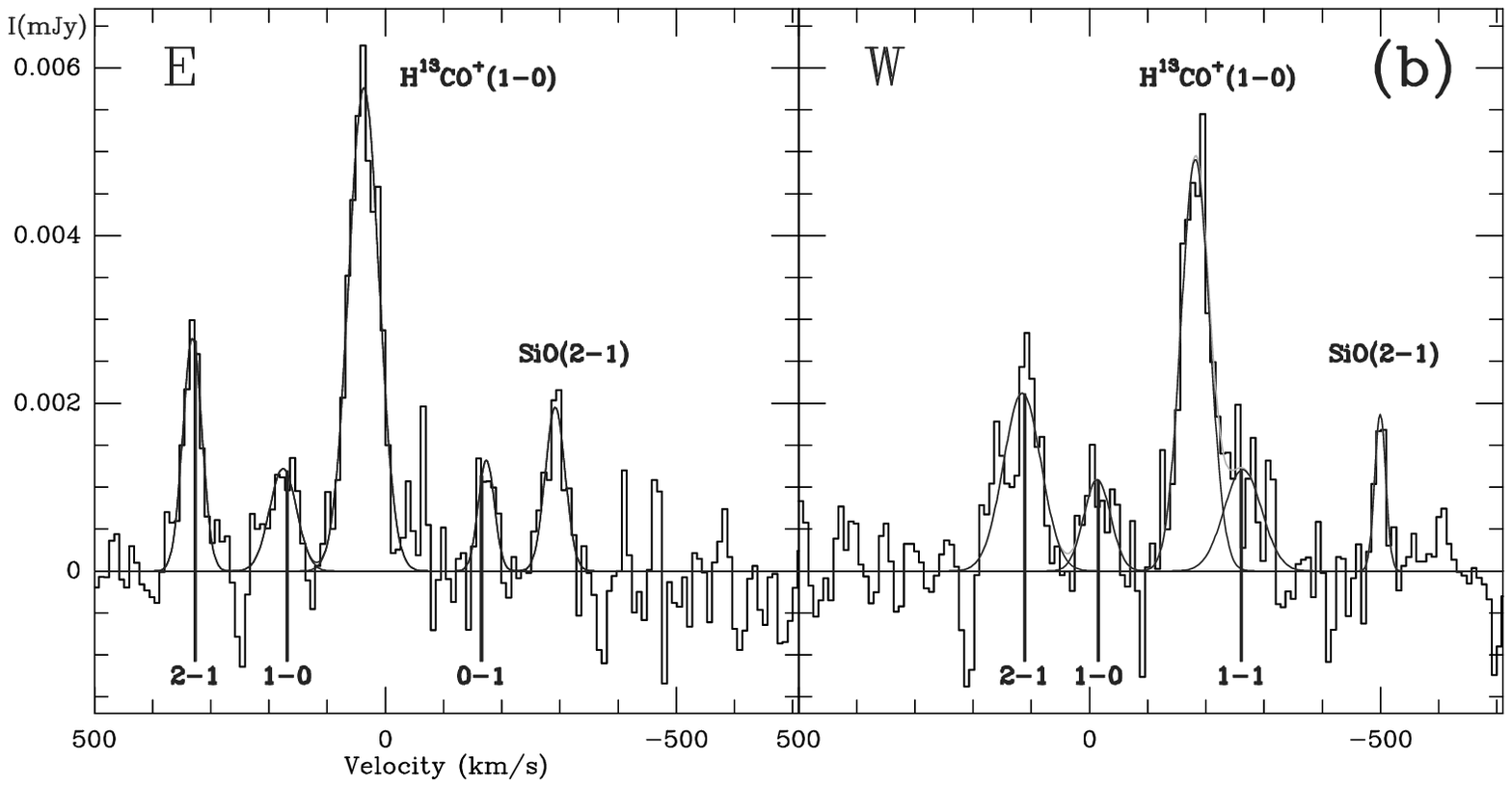}

\plotone{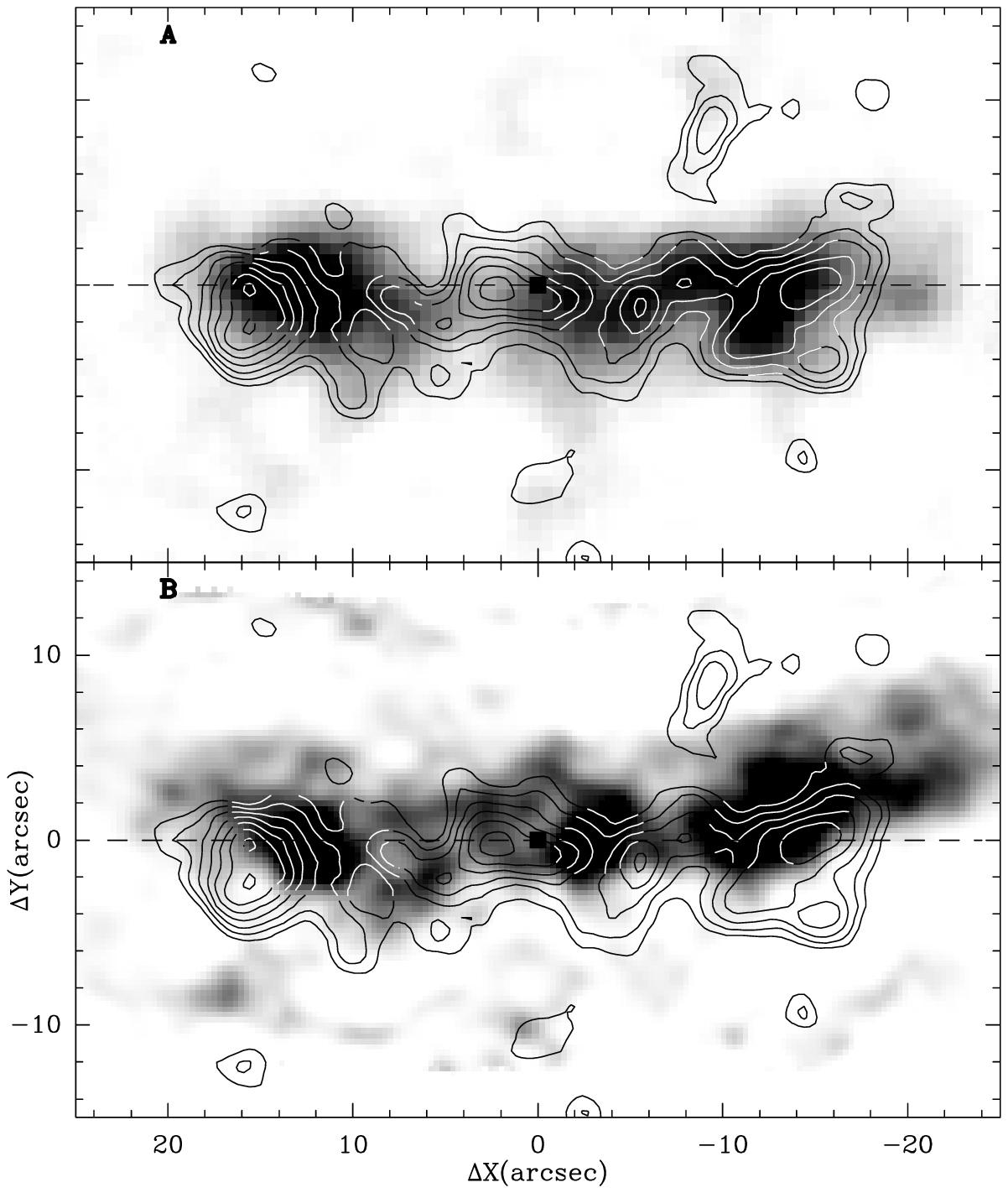}

\plotone{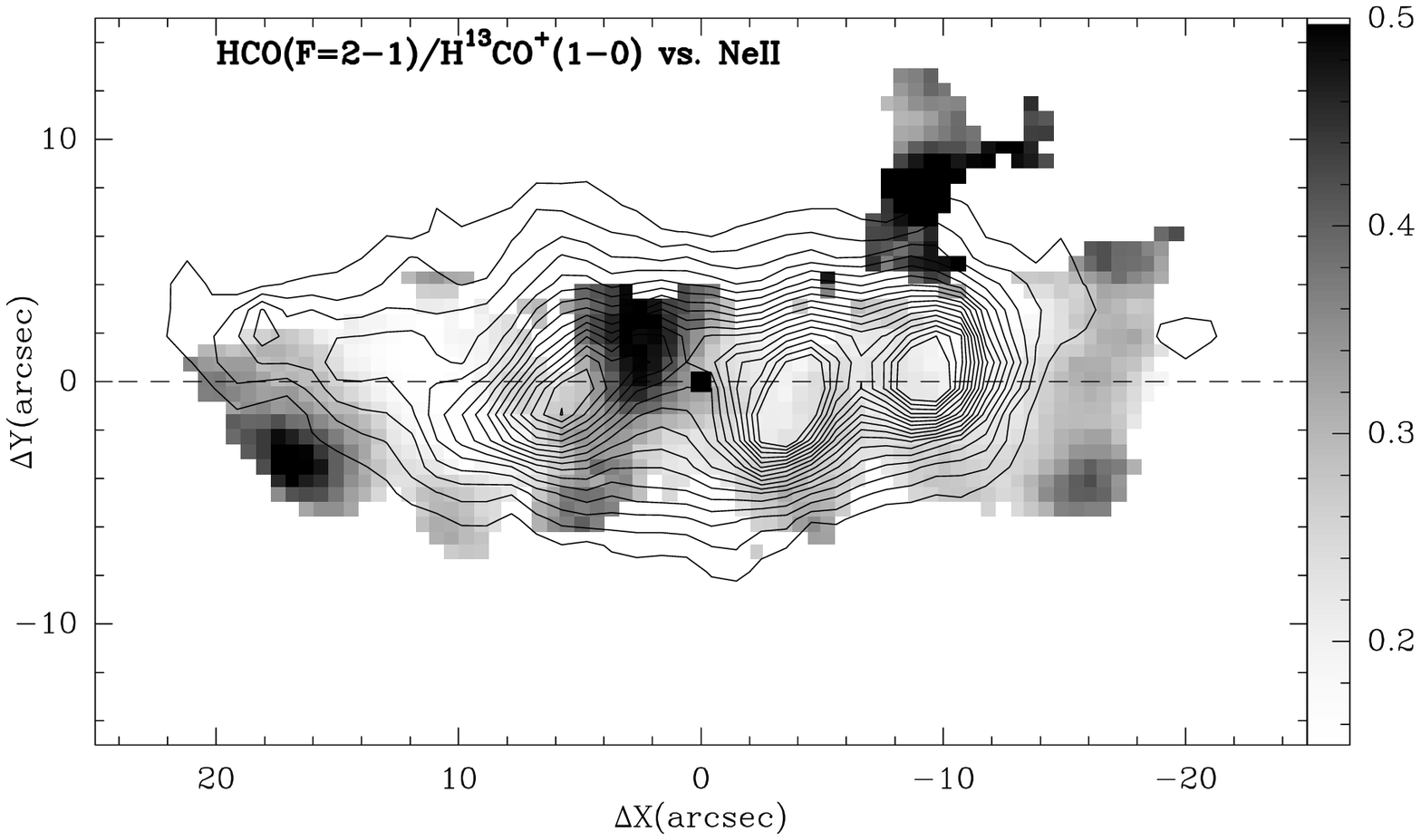}







\end{document}